\documentclass[prl,aps,twocolumn,superscriptaddress,showpacs]{revtex4}
\usepackage{graphicx}

\begin{document}

\title{Saturation induced coherence loss in coherent backscattering of
light}
\author{T. Chaneli\`{e}re}
\author{D. Wilkowski}
\email{wilkowsk@inln.cnrs.fr}
\homepage{http://www-lod.inln.cnrs.fr/} \address{Laboratoire Ondes
et D\'esordre, FRE 2302, 1361 route des Lucioles F-06560 Valbonne,
France}
\author{Y. Bidel}
\affiliation{now at Stanford University, 382 Via Pueblo Mall
CA-94305-4060 Stanford, United States}
\author{R. Kaiser}
\author{C. Miniatura}
\address{Laboratoire Ondes et D\'esordre, FRE 2302, 1361 route des
Lucioles F-06560 Valbonne, France}

\date{\today{}}

\begin{abstract}
We use coherent backscattering (CBS) of light by cold Strontium
atoms to study the mutual coherence of light waves in the multiple
scattering regime. As the probe light intensity is increased, the
atomic optical transition starts to be saturated. Nonlinearities
and inelastic scattering then occur. In our experiment, we observe
a strongly reduced enhancement factor of the coherent
backscattering cone when the intensity of the probe laser is
increased, indicating a partial loss of coherence in multiple
scattering.
\end{abstract}


\maketitle

\address{Laboratoire Ondes et D\'esordre, FRE 2302, 1361 route des Lucioles
F-06560 Valbonne, France}

\homepage{http://www-lod.inln.cnrs.fr/}

Wave coherence in the multiple scattering regime is a key
ingredient to reveal the impact of interference on wave transport
in strongly scattering media, with special interest in photonic
crystals \cite{yablonovitch}, random lasers \cite{cao}, strong and
weak localization \cite{wiersmaSL, maretUCF}. It has been shown
that wave coherence has some robust features which survive the
spatial configuration average. A clean illustration is given by
the coherent backscattering phenomenon
(CBS)\cite{wolf,wiersma95a}, a random $\emph{two-wave}$ zero path
length interferometer. CBS manifests itself as an enhanced diffuse
reflection peak in the backscattering direction. This signal is
related to the Fourier transform of the configuration-averaged
two-field correlation function (the mutual coherence) at two
space-time points at the surface of the medium \cite{vancittert}.
Hence the enhancement factor $\alpha$ (the peak to background
ratio) provides a simple measure of spatial coherence properties
of the disordered system after spatial averaging. For perfect
contrast, $\alpha$ takes its maximal value of two, a symmetry
property bearing on reciprocity \cite{reciprocity}. In recent
experiments, we have studied CBS in the elastic scattering regime
with cold atomic vapours exposed to low intensity quasi-resonant
monochromatic light. We have evidenced a loss of contrast due to
the internal structure of atoms
\cite{labeyrie99,muller02,labeyrie03} and a full contrast
restoration when non-degenerate atoms are used \cite{bidel02}.

Studying optical wave transport and localization effects with cold
atoms offer several advantages. Indeed, they act as ideal
point-like scatterers, where the scattering matrix can be fully
described with $\emph{ab initio}$ calculations and no adjustable
parameters. Moreover, the presence of sharp resonances results in
large scattering cross-sections, easily tunable by few orders of
magnitude, and large associated time scales, making resonant
scattering systems different from non-resonant multiple scattering
studied so far. If strong driving fields are used, atoms exhibit
unusual scattering properties. First the atomic susceptibility
shows up a dependence on the local field intensity. This non
linearity alters both scattering (nonlinear reduction of the
scattering cross-section) and propagation (generation of a
nonlinear refractive index for the effective medium). Second, in
addition to the usual elastic component, atoms radiate an
inelastic spectrum component. For resonantly driven atoms with a
non-degenerate groundstate, a characteristic frequency width of
this spectrum is $\Gamma$ (inverse of the excited-state lifetime)
\cite{cohen}. This inelastic spectrum is a direct consequence of
the vacuum-induced fluctuations of the driven atomic dipole. At
very strong fields, this results in the celebrated Mollow triplet
\cite{cohen, mollow}. An important question is whether the
interference effects surviving the spatial average are affected by
these non linearities and by these quantum fluctuations of the
atomic dipole. Theoretical studies investigating the impact of
$\chi ^{(2)}$ \cite{Kravtsov} and $\chi ^{(3)}$
\cite{HeiderichSkipetrov} nonlinearities predict no CBS reduction.
This seems to be supported by experimental work on CBS in gain
medium \cite{Wiersma95}. On contrary, phase fluctuations during
scattering are potential sources of a loss of coherence and a CBS
contrast reduction. These {\it dephasing} phenomena, well known in
electron transport \cite{imry}, appear to be effective as soon as
their correlation times are shorter than the wave transport time
inside the medium. With resonant scatterers like atoms, the
transport time can be very long \cite{radtrap}, of the same order
or even longer than the correlation time $\tau \simeq 1/\Gamma$ of
vaccuum-induced dipole fluctuations. One may thus expect the same
kind of decoherence mechanisms with atoms than for electrons with
an added complexity coming from the resonant character of the
scattering process. This new ingredient will induce frequency
filtering \cite{wilkowski03} and tunability in strength and shape
of the inelastic spectrum.

In this letter, we report the first experimental evidence of a
reduced CBS contrast on an optically thick cold atomic strontium
cloud when strong driving fields are used. The experimental setup
has been described elsewhere \cite{bidel02}. Typical fluorescence
measurements indicate that about $7\:10^{7}$ atoms, at a
temperature of $1mK$, are trapped in a quasi-Gaussian spherical
cloud with $rms$-size about $0.7mm$. This corresponds to a typical
atomic density at the center of the cloud about $n \simeq
10^{10}atoms/cm^{3}$. With these parameters $k \ell \approx
10^{4}$ ($k$ is the incoming wavevector and $\ell$ the light
scattering mean free path) and scattering occurs in the weak
localization regime. The maximal optical thickness achieved in our
system, as deduced from coherent transmission measurements at low
input intensity, is $b=3.5$, in reasonable agreement with the
cloud size and number of atoms.

The CBS experiment procedure uses the following time sequence.
First, the MOT is loaded during $28ms$ ($93\%$ of the duty cycle).
Then the trapping beams and the magnetic gradient are switched off
(typical falling time $1\mu s$ for the lasers and $100\mu s$ for
the magnetic field). The residual magnetic field is less than $1G$
making the Zeeman splitting small compared to the linewidth
$\Gamma$. Once the MOT is turned off, a resonant probe laser is
switched on for a short period of time. In the present study, the
probe laser parameters (intensity and frequency) are varied. The
probe pulse duration is adjusted accordingly (typically from $5$
to $70\mu s$) to keep the maximum number of absorbed photons per
atom below $400$. In this way, mechanical effects will be
negligible throughout the experiments since $400 k v_{rec} /
\Gamma \approx 0.3$, where $v_{rec}$ is the atomic recoil velocity
associated with the absorption of a single photon. Finally, most
of the atoms are recaptured during the next MOT sequence. The
collimated CBS probe laser (beam waist $2 mm$) and the response
function of our detection system yield an angular resolution well
described by a Gaussian convolution with a width $\approx
0.06mrad$, sufficiently below the typical CBS angular width
($0.3mrad$). Wave plates and polarizing optical components are
used to select the polarization of the incident probe beam and of
the detected backward fluorescence. All measurements presented in
this paper have been performed in the helicity preserving channel
($ h\Vert h$). In this channel single scattering is rejected and
an enhancement factor of two is predicted and observed at low
light intensity \cite{bidel02}. However, since the channel
isolation is not perfect in the experiment, single scattering will
spoil the signal. This happens preferentially at low optical
thickness because single scattering has the largest contribution
to the total backscattered signal. For our experiments this effect
leads to an enhancement factor reduction of few percents.

The principle of the CBS detection scheme has been described in
ref. \cite{labeyrie00}. The far-field backward fluorescence signal
is collected on a cooled CCD Camera. A mechanical chopper is
placed between the MOT and the CCD. It is synchronized with the
full time sequence in order to close the detection path when the
MOT is operating and to open it when the probe beam is switched
on. The total exposure time required for good signal-to-noise
ratio is of the order of a few seconds. Once the full signal is
collected, the acquisition is repeated during the same amount of
time, maintaining the MOT magnetic gradient off, to obtain the
background signal. This stray signal, corresponding to $15\%$ of
the total signal, is then subtracted to get the CBS signal. A 2D
fitting procedure is then used to extract the main CBS cone
parameters, its width and its enhancement factor. The theoretical
shape of the CBS cone implemented in the fitting procedure is
given by a Monte-Carlo simulation performed at low saturation but
taking into account the Gaussian distribution of atoms in a cloud
\cite {bidel02}. Increasing the probe beam intensity did not
reveal any significant change in the shape of the CBS cone, at
least in the range of parameters used in our experiment. This is
the reason why we treat all the data with the same cone shape. The
finite angular resolution of our apparatus has been taken into
account by convolving the preceding theoretical CBS cone shape by
a Gaussian having the measured apparatus angular width.

\begin{figure} [tbp]
\begin{center}
\includegraphics[scale=1.5]{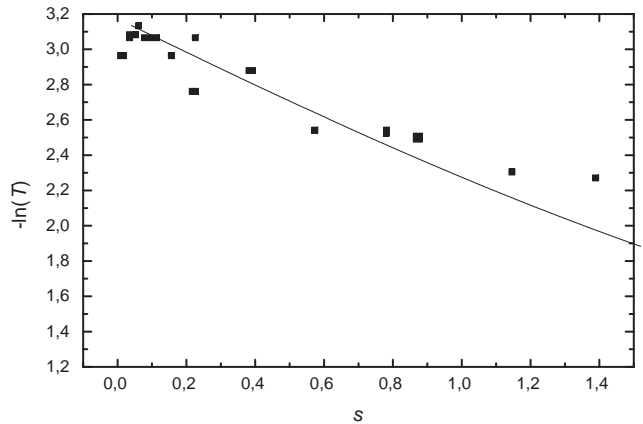}
\end{center}
\caption{Resonant ($\delta=0$) coherent transmission $T$ along a
diameter of the cold strontium could as a function of the
saturation parameter $s$. The squares are the experimental values
and the solid line correspond to the theoretical Lambert-Beer
prediction taking into account the nonlinear reduction of the
scattering cross-section.} \label{trans}
\end{figure}

Beyond the complexity of the situation under consideration
(multiple scattering with nonlinear and inelastic scatterers), one
has to deal also with nonuniform scattering properties. Indeed,
even in an homogeneous slab geometry, the local intensity is not
constant, as the incident coherent beam is attenuated when
penetrating into the medium. Hence the atoms located deeper inside
the medium will not be saturated in the same way as the atoms on
the front part of the sample. Thus the saturation, and hence the
scattering cross-section, will not be constant along a given
multiple scattering path. The importance of the spatial variation
of the saturation parameter can be estimated by looking at the
attenuation of the coherent beam. In Fig. \ref{trans} we report
the measured transmission and we compare it with the Lambert-Beer
theoretical prediction taking into account the non-linear
reduction of the cross-section. If one assumes that the local
atomic saturation is dominated by the incident field and not by
the scattered field, this theoretical curve is obtained by solving
the following equation:
\begin{equation}
\frac{ds}{dz}=\frac{s}{(1+s)\ell}
\label{atte}
\end{equation}
Here $s$ is the saturation parameter defined as:
\begin{equation}
s=\frac{I/I_{sat}}{1+(2 \delta /\Gamma )^{2}} \label{sat}
\end{equation}
where $I_{sat}$ is the saturation intensity ($I_{sat}=42mW/cm^{2}$
for Sr) and where $\delta $ is the laser detuning with respect to
the atomic transition. The factor $1/(1+s)$ features the non-linear reduction of
the scattering efficiency and one gets the normal Lambert-Beer law when
$s \to 0$. The low-intensity
scattering mean free path $\ell$ reads :
\begin{equation}
\ell (\delta) = \frac{1}{n \sigma (\delta)}= \frac{(1+(2 \delta /\Gamma )^{2}}{n \sigma_0}
\label{mfp}
\end{equation}
with the resonant low-intensity scattering cross-section $\sigma_0
= 3 \lambda^2 / 2 \pi$. The excellent agreement with the measured
attenuation prooves that saturation plays a role in our
experimental conditions (since otherwise the transmission would
not depend on $s$) and that the local atomic saturation is indeed
dominated by the incident field.

\begin{figure}[tbp]
\begin{center}
\includegraphics[scale=1.25]{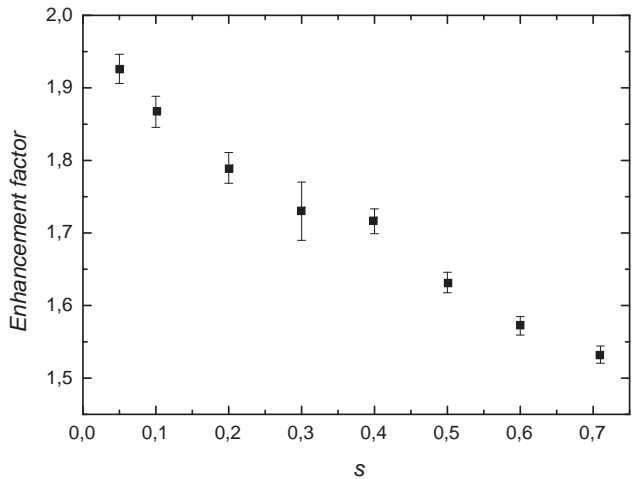}
\end{center}
\caption{Resonant ($\delta=0$) CBS enhancement factor as a
function of the incident saturation parameter $s$. The coherent
transmission value is kept fixed to $T=0.085$.} \label{ConeRes}
\end{figure}

Fig. \ref{ConeRes} shows the dependence of the CBS enhancement
factor as a function of the incident saturation parameter $s$,
with the probe maintained at resonance ($\delta =0$). In principle
one would like to vary the saturation parameter without modifying
the relative weight of the various scattering orders involved in
the CBS signal. This however proves to be difficult to assure
because of the modification of the atom scattering properties when
$s$ increases. In order to minimize any effect relating to a
modification of the distribution of scattering orders, we tried to
keep the coherent beam profile throughout the sample as constant
as possible. This is achieved by adjusting, for each value of $s$,
the total number of cold atoms in the cloud in order to maintain
the coherent transmission $T$ as constant as possible ($T \simeq
0.085$ in the data shown in Fig. \ref{ConeRes}). As shown in
Fig.\ref{ConeRes} we observe an enhancement factor of $1.93\pm
0.02$ at low saturation. The small systematic reduction of $\alpha
$ compared to the expected value of two is in agreement with the
presence of residual single scattering in the \textit{forbidden}
$h\Vert h$ polarization channel, as we previously discussed. The
most striking feature is the rapid quasi-linear decrease of the
enhancement factor as $s$ is increased. The slope derived from a
$\textit{rms}$-procedure is $(\delta \alpha /\delta s)\approx
-0.6$. As the transmission is kept fixed, we estimate numerically
the fraction of single scattering to increase by less than $10\%$
when the saturation parameter is increased up to $s=0.8$. The
associated reduction of the enhancement factor should be of the
order of $1\%$, negligible compared to the observed reduction.
Thus the CBS reduction comes from the multiple scattering signal.

\begin{figure}[tbp]
\begin{center}
\includegraphics[scale=1.5]{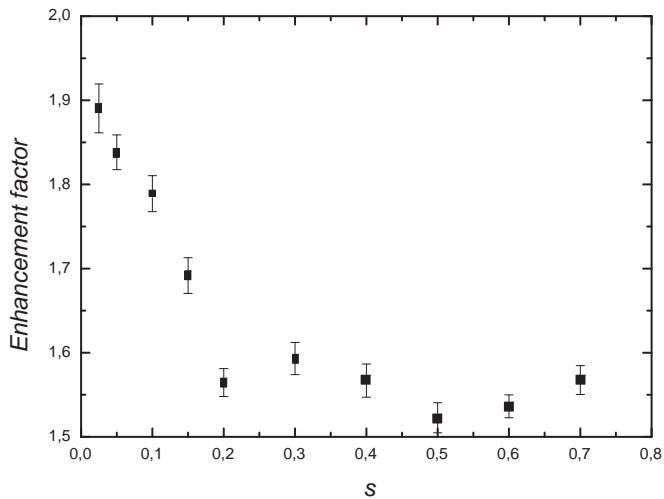}
\end{center}
\caption{Off-resonant ($\delta=\Gamma/2$) CBS enhancement factor
as a function of the incident saturation parameter $s$. The
coherent transmission value is kept fixed to $T=0.19$. Compared to
the resonant case, the overall behavior is different with a
stronger dicrease at low $s$.} \label{ConeDet}
\end{figure}

In order to see to what extent the resonance affects the coherence
properties probed by CBS, we performed another experiment at
$\delta =\Gamma /2$. The  same experimental procedure has been
used with a transmission now at $T=0.19$. As shown in Fig.
\ref{ConeDet} a different general behavior is observed. First, at
low intensity, the linear decreasing is faster since $(\delta
\alpha /\delta s)\approx -1.8$. Second, for larger saturation
parameters ($0.3<s<0.8$) the decrease is then slowed down. The two
sets of data in Fig.\ref{ConeRes} and \ref {ConeDet} are obtained
with a different transmission value, but other studies show that
the enhancement factor does not sensitively depend on the
transmission value. So, if we compare these data, it shows that
$s$ is not the only relevant parameter in our experiment. Indeed,
the exact shape of the inelastic spectrum also depends on the
detuning $\delta$. In particular, for the detuned case, part of
the inelastic spectrum will overlap the atomic resonance. This
resonant inelastic light will thus be scattered again more
efficiently then the off-resonant elastic part. This effect is
{\it e.g.} responsible for an increase of the MOT volume in the
multiple scattering regime \cite{sesko91}. Finally in our
experiment, the ratio inelastic versus elastic multiple scattered
light may change with the detuning. We may then conclude that the
CBS reduction is due to the inelastic spectrum, but one has also
to keep in mind that the dispersive aspect of the atomic response
to a driving field implies that non linearities at propagation and
at scattering are drastically different for on-resonant and
detuned excitations.

In summary, we observed that coherent backscattering is strongly
reduced when the atomic transition is saturated, a signature of
the wave coherence loss in the multiple scattering regime. We
speculate that the origin of this reduction is due to the
inelastic scattering of the light by the atoms, {\it i.e.} from
the coupling of the atomic dipole to the vacuum modes of the field
rather than to the non-linear response of the atomic dipole. The
different behavior of the CBS enhancement factor for resonant and
off-resonant excitations (see Fig. \ref{ConeRes} and
\ref{ConeDet}) indicates that the saturation parameter $s$ alone
does not allow for an universal scaling. It is important to
realize that, in the quest of strong localization of light in
disordered medium, large local build-up of the intensity can occur
in the localized states. If the localization length is of the
order of few optical wavelength, a single resonant photon could in
principle saturate the atoms located in that region, in analogy
with cavity QED effects \cite{kimble}. The observations reported
in this letter are thus important for the study of strong
localization of light in atomic vapours.

The authors thank D. Delande and G. Labeyrie for fruitful
discussions. This research is financially supported by the CNRS,
the PACA region and the BNM.

----------------------------------------------------------------

\begin{thebibliography}{99}

\bibitem{yablonovitch}  E. Yablonovitch, Phys. Rev. Lett. \textbf{58}, 2059
(1987).

\bibitem{cao} H. Cao \textit{et al.}, Phys. Rev. Lett. \textbf{84}, 5584
(2000); X. Jiang and C. M. Soukoulis, Phys. Rev. Lett.
\textbf{85}, 70 (2000); P. Sebbah and C. Vanneste, Phys. Rev. B
\textbf{66}, 144202 (2002).

\bibitem{wiersmaSL} D.S. Wiersma, P. Bartolini, A. Lagendijk and R. Righini,
Nature \textbf{390}, 671 (1997).

\bibitem{maretUCF} F. Scheffold and G. Maret, Phys. Rev. Lett. \textbf{81}, 5800
(1998).

\bibitem{wolf}  P.E. Wolf and G. Maret, Phys. Rev. Lett. \textbf{55}, 2696
(1985); M.P. van Albada and A. Lagendijk, Phys. Rev. Lett.
\textbf{55}, 2692 (1985).

\bibitem{wiersma95a}  D.S. Wiersma, M.P. van Albada, B.A. van Tiggelen and A. Lagendijk, Phys.
Rev. Lett. \textbf{74}, 4193 (1995).

\bibitem{vancittert} This formulation of CBS is a
consequence of the van Cittert-Zernike Theorem. See for example:
M. Born and E. Wolf, {\it Principes of Optics}, Cambridge
University Press, Cambridge (1999).

\bibitem {reciprocity} B. van Tiggelen and R. Maynard {\it in Waves in random and other complex media}, R.
Burridge, G. Papanicolaou and L. Pastur (eds.) \textbf{96},
Springer (1997), p 247.

\bibitem{labeyrie99}  G. Labeyrie, F. de Tomasi, J.C. Bernard,
C.A. M\"{u}ller, C. Miniatura and R. Kaiser , Phys. Rev. Lett.
\textbf{83}, 5266 (1999).

\bibitem{muller02} C. A. M\"{u}ller and Ch. Miniatura, J. Phys. A: Math. Gen.
35, 10163 (2002).

\bibitem{labeyrie03}  G. Labeyrie, D. Delande, C. A. M\"{u}ller, C. Miniatura, R.
Kaiser, Phys. Rev. A \textbf{67}, 033814 (2003); G. Labeyrie, D.
Delande, C. A. M\"{u}ller, C. Miniatura, R. Kaiser, Europhys.
Lett. \textbf{61}, 327 (2003).

\bibitem{bidel02}  Y. Bidel, B. Klappauf, J.C. Bernard, D. Delande, G.
Labeyrie, C. Miniatura, D. Wilkowski, R. Kaiser, Phys. Rev. Lett.
\textbf{88}, 203902 (2002).

\bibitem{cohen}  C. Cohen-Tannoudji, J. Dupont-Roc, G. Grynberg, \textit{Atom-Photon Interactions}, Wiley
(1992).

\bibitem{mollow}  B. Mollow, Phys. Rev. \textbf{188}, 1969 (1969).

\bibitem{Kravtsov}  V. Agranovich and V. Kravtsov, Phys. Rev. B \textbf{43 }%
, 13691 (1991).

\bibitem{HeiderichSkipetrov}  A. Heiderich, R. Maynard and B. van Tiggelen,
Opt. Comm. \textbf{115}, 392 (1995); S.E. Skipetrov and R. Maynard
\textit{in Wave Scattering in Complex Media: from theory to
applications}, NATO Science Series II \textbf{107}, eds. B.A. van
Tiggelen and S.E. Skipetrov, Kluwer, Dordrecht (2003), p. 75.

\bibitem{Wiersma95}  D.S. Wiersma, M.P. van Albada and A. Lagendijk, Phys.
Rev. Lett. \textbf{75}, 1739 (1995).

\bibitem{imry} A. Stern, Y. Aharonov, and Y. Imry, Phys. Rev. A \textbf{41}, 3436
(1990).

\bibitem{radtrap} G. Labeyrie, E. Vaujour, C.A. M\"{u}ller, D. Delande, C. Miniatura,
D. Wilkowski and R. Kaiser,  submitted to Phys. Rev. Lett. (2003).

\bibitem{wilkowski03}  D. Wilkowski, Y. Bidel, T. Chaneli\`{e}re, R. Kaiser,
B. Klappauf, C.A. M\"{u}ller and Christian Miniatura Phys. B
\textbf{328}, 157 (2003).

\bibitem{labeyrie00}  G. Labeyrie, C. M\"{u}ller, D. Wiersma, C. Miniatura
and R. Kaiser, J. Opt. B : Quantum Semiclass. Opt. \textbf{2 },
672 (2000).


\bibitem{sesko91}  D. Sesko, T. Walker and C. Wieman, JOSA B \textbf{8}, 946
(1991).

\bibitem{kimble}  Daniel Kleppner, Phys. Rev. Lett. \textbf{47}, 233 (1981)


\end{thebibliography}
\end{document}